# THEORY AND TECHNIQUES FOR VIBRATION-INDUCED CONDUCTIVITY FLUCTUATION TESTING OF SOILS

HUNG-CHIH CHANG, LASZLO B. KISH

*Department of Electrical and Computer Engineering, Texas A&M University, College Station, TX 77843-3128, USA*

A. KISHNÉ AND C. MORGAN

*Department of Soil and Crop Sciences, Texas A&M University, College Station, TX 77843-2474, USA*

CHIMAN KWAN

*Signal Processing, Inc., 13619 Valley Oak Circle, Rockville, MD 20850, USA*

**Abstract**

First we present and theoretically analyze the phenomenological physical picture behind Vibration-Induced Conductivity Fluctuations. We identify the relevant tensors characterizing the electromechanical response against the vibrations for both longitudinal and transversal responses. We analyze the conductivity response with acceleration type vibrations and a new scheme, measurements with more advantageous compression type vibrations that are first introduced here. Compression vibrations provide sideband spectral lines shifted by the frequency of the vibration instead of its second harmonics; moreover the application of this method is less problematic with loose electrodes. Concerning geometry and electrodes, the large measurement errors in earlier experiment indicated electrode effects which justify using four-electrode type measurements. We propose and analyze new arrangements for the longitudinal and transversal measurements with both compression vibration and acceleration vibration for laboratory and field conditions.

## 1. Introduction: principle and published experiments

Bulk soil electrical conductivity is influenced mainly by water and dissolved salts in the pore phase in saturated moisture conditions, but the solid phase also contributes to current flow through direct and continuous contact with one another, and via exchangeable actions on the surface of clay minerals [1, 2]. As the soil dries, the relative contribution of solids, particularly hydrated clay minerals, increases with approaching the current percolation threshold, and it is also affected by the spatial distribution of particles [3-5]. Bulk or apparent electrical conductivity ($EC_a$) sensors were designed initially to quantify salt concentration across large agricultural fields. However at low salt concentration in more humid regions, they have been useful in identifying clay content, soil moisture, and bulk density [6]. Electromagnetic field techniques include electromagnetic induction and ground penetrating radar as proximal sensors, and electrical conductivity or resistivity, capacitance, and time-domain reflectometry as sensors that are in direct contact with the soil [7, 8]. These measurements are advantageous because their application is fast and can be conducted in-situ; however, high salt concentration limits their applicability [2, 8].

A new technology, vibration-induced conductivity fluctuation (*VICOF*), measures the resistance variation modulated by vibration in direct contact with the soil, and provides new information on soil structure and porosity in addition to electrical conductivity [9,

*Introducing compression VICOF and the phenomenological theory of compression and acceleration VICOF.*

10]. The advantage of *VICOF* is the reduced effect of salinity on the interpretation of $EC_a$ applied to other soil properties.

The measurement setup is based on the AC conductivity circuit shown in Figures 1 and 2 [9, 10]. The evaluation formula for the AC resistance $R_s$ of the soil sample is [9]:

$$R_s = R_1 \frac{U_{2,1}}{U_1 - U_{2,1}} \quad , \tag{1}$$

where $U_1$ is the driving sinusoidal voltage amplitude with frequency $f_1$ and $U_{2,1}$ is the voltage measured on the soil at frequency $f_1$.

Small periodic vibration at frequency $f_2$ ($<< f_1$) of the soil will generate fluctuations $dR_s$ in the soil resistance. The normalized $dR_s/R_s$ can be determined from the measurement with the following equation [9]:

$$\frac{dR_s}{R_s} = 2\frac{U_{2,2}}{U_{2,1}}(1 + \frac{U_{2,1}}{U_1 - U_{2,1}}) \quad , \tag{2}$$

where $U_{2,2}$ is the voltage amplitude measured at any of the combination frequencies $f_1 + 2f_2$ or $f_1 - 2f_2$.

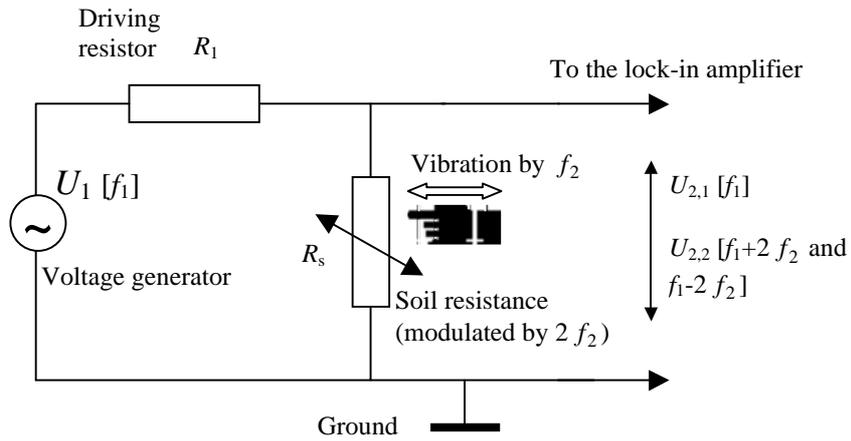

**Figure 1.** Measurement Circuitry.

It is important to note that the second term in the parenthesis of the right-hand side of Equation 2 will be zero if, instead of the voltage generator and corresponding $R_1$, we use an AC current generator to drive current through the soil sample.

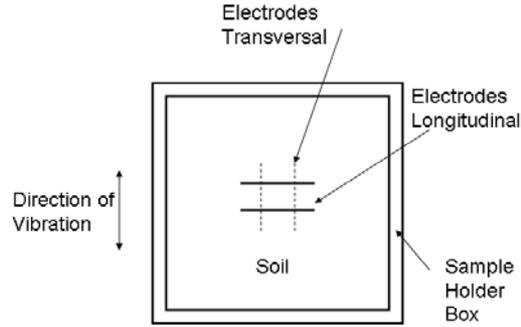

**Figure 2.** Setup arrangement of the measurement. The sample holder box is made of insulator.

The reason is that the current generator corresponds to infinite $R_1$ and infinite $U_1$. Thus, for current generator driving:

$$\frac{dR_s}{R_s} = 2\frac{U_{2,2}}{U_{2,1}} \qquad (3)$$

(Note, the factor of 2 stems from the fact that the amplitudes at the combination frequencies are half of the amplitudes of the corresponding DC arrangement.)

To further model the applied vibration and soil resistance variation, the relationship between the applied force, stress, strain and resistance are presented in the following sections.

2.  **The electromechanical response and its tensors for horizontal acceleration-based vibration**

There are two kinds of *VICOF*, vibration based on acceleration and compression; *AVICOF* and *CVICOF*, respectively. The horizontal vibration acceleration *VICOF* model is shown in Figure 3. The test soil sample is a cube with length $L_x$ in x-direction and $L_y$ in y-direction. The resistance measurement point is in the center of the sample. Suppose, we apply a periodic vibration

$$A_x(t) = A_0 \sin \omega t \qquad (4)$$

in the direction *x*, where $A_0$ is the vibration amplitude and $\omega$ is the angular frequency of the vibration. The acceleration of the soil can be expressed as

*Introducing compression VICOF and the phenomenological theory of compression and acceleration VICOF.*

$$a_x(t) = -\omega^2 A_0 \sin \omega t \quad . \tag{5}$$

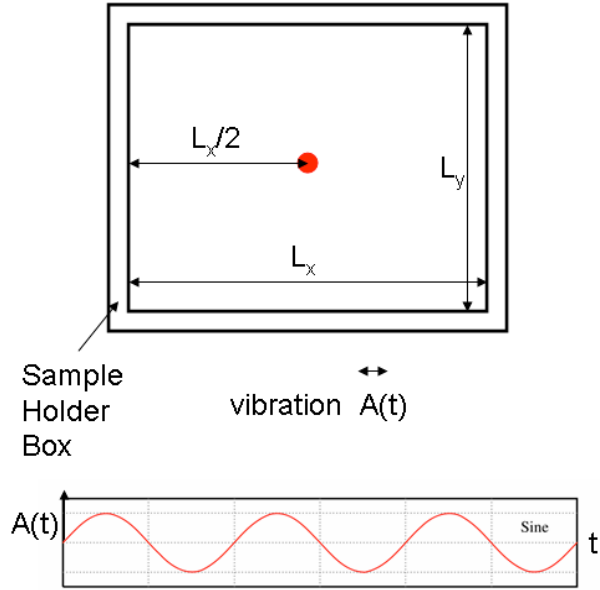

**Figure 3.** Horizontal acceleration-based *VICOF* model

As a consequence, the acceleration of the soil induces a stress $\sigma$ (force per unit area) and strain $\epsilon$ (relative deformation) in the soil

$$\sigma_x(t) = \rho \cdot \omega^2 A_0 \mid \sin \omega t \mid \cdot f_\sigma(L) \quad , \tag{6}$$

$$\varepsilon_x(t) = \rho \cdot \omega^2 A_0 \mid \sin \omega t \mid \cdot f_\varepsilon(L) \quad , \tag{7}$$

where $\rho$ is the bulk density ($kg\ m^{-3}$) of the soil; $\sigma_x$ and $\epsilon_x$ are the components of stress and strain in the x direction, respectively; and $f_\sigma(L)$ and $f_\varepsilon(L)$ are functions of the location *L*.

The stress of the soil can generate the strain of the soil through the Hook's law. For a three-dimensional state of stress within the linear elastic range, each of the six stress components (three normal stress components *x*, *y*, and *z* and three shear stress components *xy*, *xz*, and *yz*) is expressed as a linear function of the six components of strain, and vice versa. The stress-strain relationship is shown in Equations (8) and (9), where [*S*] is the Compliance Matrix.

$$[\varepsilon] = [S][\sigma] \Rightarrow [\varepsilon] \propto [\sigma]] \qquad (8)$$

$$\begin{bmatrix} \varepsilon_{XX} \\ \varepsilon_{YY} \\ \varepsilon_{ZZ} \\ \varepsilon_{YZ} \\ \varepsilon_{ZX} \\ \varepsilon_{XY} \end{bmatrix} = \begin{bmatrix} S_{11} & S_{12} & S_{13} & S_{14} & S_{15} & S_{16} \\ S_{21} & S_{22} & S_{23} & S_{24} & S_{25} & S_{26} \\ S_{31} & S_{32} & S_{33} & S_{34} & S_{35} & S_{36} \\ S_{41} & S_{42} & S_{43} & S_{44} & S_{45} & S_{46} \\ S_{51} & S_{52} & S_{53} & S_{54} & S_{55} & S_{56} \\ S_{61} & S_{62} & S_{63} & S_{64} & S_{65} & S_{66} \end{bmatrix} \begin{bmatrix} \sigma_{XX} \\ \sigma_{YY} \\ \sigma_{ZZ} \\ \sigma_{YZ} \\ \sigma_{ZX} \\ \sigma_{XY} \end{bmatrix}. \qquad (9)$$

Further, the stress generates the connection or disconnection of the soil particles. The probability matrix of the bondage connectivity $[P]$ is induced by stress $[P_{stress}]$.

$$[P] = [P_{stress}] = [\Phi][\varepsilon] \qquad (10)$$

Lastly, the bondage connect probability $[P]$ can transfer to the resistance $[R]$ through the equation (11) and (12), where $[T]$ is the transfer matrix,

$$[R] = [T][P], \qquad (11)$$

$$[R] = \begin{bmatrix} R_x \\ R_y \\ R_y \end{bmatrix} = [T][P] = \begin{bmatrix} T_{xx} & T_{xy} & T_{xz} \\ T_{xy} & T_{yy} & T_{yz} \\ T_{zx} & T_{zy} & T_{zz} \end{bmatrix} \begin{bmatrix} P_x \\ P_y \\ P_y \end{bmatrix}. \qquad (12)$$

Similarly, the variation of the connectivity $[\Delta P]$ can transfer to the variation of the resistance $[\Delta R]$:

$$[\Delta R] = [U(P)][\Delta P], \qquad (13)$$

$$[\Delta R] = \begin{bmatrix} \Delta R_x \\ \Delta R_y \\ \Delta R_y \end{bmatrix} = [U(P)][\Delta P] = \begin{bmatrix} U_{xx}(P) & U_{xy}(P) & U_{xz}(P) \\ U_{xy}(P) & U_{yy}(P) & U_{yz}(P) \\ U_{zx}(P) & U_{zy}(P) & U_{zz}(P) \end{bmatrix} \begin{bmatrix} \Delta P_x \\ \Delta P_y \\ \Delta P_y \end{bmatrix}. \qquad (14)$$

*Introducing compression VICOF and the phenomenological theory of compression and acceleration VICOF.*

## 3. Four-electrode arrangements and their proposed schemes for field studies with pressure vibrations

### 3.1. *The Wenner four-electrode measurement: Point contacts.*

3.1.1. *Derivation of the resistivity and the VICOF response*

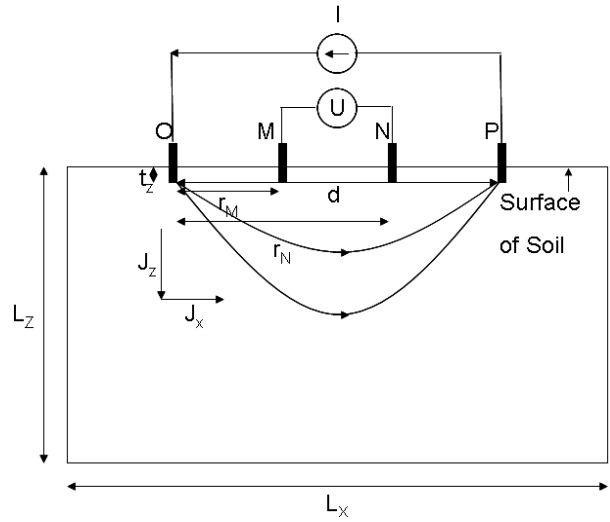

**Figure 4.** Scheme of the Wenner four-electrode (four-point) measurement.

The first electrode setup to analyze for *VICOF* response is for the Wenner four-electrode measurement method [11] as shown in Figure 4. For the validity of the Wenner method the contacts must be effectively point contacts which mean that the contact diameter and length $t_z$ are much less than the minimal spacing between the contacts and the sample size must be much greater than these.

The resistivity and the *VICOF* response for a generic four-electrode-contact arrangement (contacts aligned) will be derived by the "salami" method (summing up resistance of infinitesimally slices between equipotential surfaces) and the superposition theorem. The resistivity results for the widely used symmetric case, where the distances between the electrodes are the same, are already known as the Wenner method; however, we present a derivation with more general conditions for completeness which, in the case of uniform spacing of electrodes, reproduces the Wenner results.

We measure the voltage ($U_{mn}$) between electrodes *M* and *N* while applying a current source (I) between electrodes *O* and *P*. The distance between *O*, *M*, *N* and *P* are $\overline{OM} = r_M$, $\overline{ON} = \overline{OM} + \overline{MN} = r_N$ and $\overline{NP}$. It is important that the diameter and the length of the electrodes must be much less than $r_M$, which means that this is a point contact arrangement. Therefore, in this test set, the electrical current density has significant components in all the downward directions below the soil surface and that also means that the longitudinal and transversal *VICOF* components are mixed for any vibration direction and arrangement.

We assume first that sample size is infinite. Assume that the current enters through a single point contact with radius $r_E$ (diameter $2r_E$) and the other electrode (ground) is an infinitely large semi-sphere with center on the point contact. Using the "salami method" between two semi-spheres of radius $r$ and $r+dr$ centered on the point electrode, the resistance contribution of this infinitesimally thin layer (of thickness $dr$) to the total resistance of the point contact is given as:

$$dR = \frac{\rho \cdot dr}{2\pi r^2} \quad . \tag{15}$$

The voltage drop between the two equipotential surfaces is

$$dU = \frac{I\rho \cdot dr}{2\pi r^2} \quad . \tag{16}$$

Thus the voltage induced by *O* between the equipotential semi-sphere surface of radius $r$ and the ground is:

$$U(r) = \int_r^\infty \frac{I\rho}{2\pi r^2} dr = \frac{I\rho}{2\pi r} \quad . \tag{17}$$

Now we consider the arrangement in Figure 4. Using the superposition theorem and the fact that the current direction at the two electrodes is opposite, we can easily calculate the voltage difference between the pickup electrodes at points *M* and *N*.

The voltage between point *M* and *N* induced by the current through electrode *O* is

$$U_{MN,O} = \int_{r_M}^{r_N} \frac{I\rho}{2\pi r^2} dr = \frac{I\rho}{2\pi}\left(\frac{1}{r_M} - \frac{1}{r_N}\right) \quad . \tag{18}$$

Similarly, the voltage between point *M* and *N* induced by the negative current through electrode *P* is

$$U_{MN,P} = -\int_{d-r_M}^{d-r_N} \frac{I\rho}{2\pi r^2} dr = \frac{I\rho}{2\pi}\left(\frac{1}{d-r_N} - \frac{1}{d-r_M}\right) \quad . \tag{19}$$

The negative sign is for the inverse current direction (current flowing in via *O* and flowing out through *P*). Therefore the total voltage drop between electrodes *M* and *N* is the superposition of $U_{MN,O}$ and $U_{MN,P}$:

*Introducing compression VICOF and the phenomenological theory of compression and acceleration VICOF.*

$$U_{MN} = \frac{I\rho}{2\pi}\left(\frac{1}{r_M} - \frac{1}{r_N} + \frac{1}{d-r_N} - \frac{1}{d-r_M}\right) \quad . \tag{20}$$

The resistivity of the soil is

$$\rho = \frac{2\pi U_{MN}}{\left(I\frac{1}{r_M} - \frac{1}{r_N} + \frac{1}{d-r_N} - \frac{1}{d-r_M}\right)} \quad . \tag{21}$$

Equation 25 is the general result for arbitrary point electrode distances.

In the Wenner electrode arrangement [11], it is supposed that the electrodes have equal spacing *r*, that is:

$$\overline{OM} = \overline{MN} = \overline{NP} = r \; ; \text{ therefore, } d = 3r. \tag{22}$$

Thus the total voltage between electrodes is given as

$$U_{MN} = \frac{I\rho}{2\pi r} \quad . \tag{23}$$

The resistivity of the soil is

$$\rho = \frac{2\pi r U_{MN}}{I} \quad . \tag{24}$$

Equation 28 is the same result given in Moorey's paper [11].

The normalized resistivity fluctuation at a DC-voltage based *VICOF* response would be:

$$\frac{\Delta \rho_v}{\rho} = \frac{\Delta U_{MN}}{U_{MN}} \quad . \tag{25}$$

However in the *VICOF* method, we used an AC method to avoid polarization artifacts around the electrodes. Thus we must take into the account that, at a given sideband (combination frequency), the relative amplitude is half of the relative conductance fluctuation response, see Equation 3. That yields:

$$\frac{\Delta \rho_v}{\rho} = \frac{2U_{MN}(f_c)}{U_{MN}(f_1)} \quad . \tag{26}$$

where $U_{MN}(f_1)$ is the voltage measured at frequency $f_1$ and $U_{MN}(f_c)$ is the voltage measured at any of the combination frequencies $f_c \equiv f_1 \pm f_2$.

### 3.1.2. *The disadvantage of the Wenner four-electrode arrangement for VICOF applications*

The normalized resistivity fluctuation $\Delta\rho_V / \rho$ is the most important quantity of *VICOF* measurements. However, the transversal and longitudinal VICOF components are usually different and it is desirable to measure them separately whenever it is possible; just to measure one of them without mixing in the other.

However, the Wenner method is mixing the transversal and longitudinal components due to the various directions of the current density. Due to this situation, the transversal and longitudinal components cannot be separated. Therefore, we can conclude that the Wenner electrode arrangement is not advantageous for applying it in the *VICOF* studies, even though it can be used as a quick assessment tool if needed.

We note that the line-contact method, when used together with the vertical compression *VICOF* method (see below in Sections 3.2 and 4), as well as the plate-contact method with any VICOF arrangements (see below in Section 3.3), are free of the disadvantage of mixing transversal and longitudinal components.

### 3.2. *Four-electrode measurement with line-contacts*

This arrangement is similar to the Wenner four-electrode method except that the contacts are not point contacts but effectively two-dimensional line-contacts. That means that the electrode length is much greater than the electrode distance at field measurements and, at lab measurements, it is as long as the box size. Under these conditions, this arrangement will still mix the transversal and longitudinal *VICOF* components when the vibration is in the horizontal direction. However, with a new *VICOF* arrangement, which we call the vertical compression *VICOF* and vertical acceleration *VICOF*, see below in Section 4, a clear transversal response can be produced because the current density lines are horizontal in the box-based lab setup and dominantly horizontal in the field setup. Therefore, this arrangement is practically important because of the convenience of installing the rod electrode system.

The key is to use this arrangement to practically eliminate the vertical (downward) component of current density in the four-electrode measurement. The vertical current density components will arise only from the end of the current-driven rod electrodes and, if the length of the rod is much longer than its diameter, the dominant current term will be horizontal.

Moreover, the most ideal case can be realized during a lab measurement by selecting the length of the rod electrode so that it goes through the whole depth of the soil and the soil is in an insulating box, see Figures 5 and 6. In this particular case, all current density lines will be horizontal. In this case, the current electrodes *O* and *P* can be as long as the voltage electrode rods *M* and *N* because due to the limitation presented by the bottom of the insulating box, there are no vertical current density lines and neither can such lines arise at the top of the electrodes. Therefore, the voltage electrodes only sample the voltage and do not short circuit points with different potentials therefore they do not disturb the current density distribution provided by the voltage electrodes.

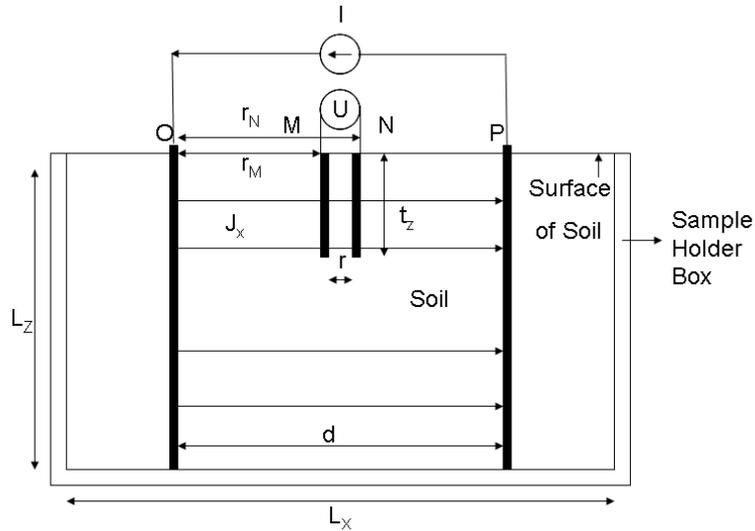

**Figure 5.** Side view of setup arrangement of the four-electrode measurement with line-contacts in a box of insulator.

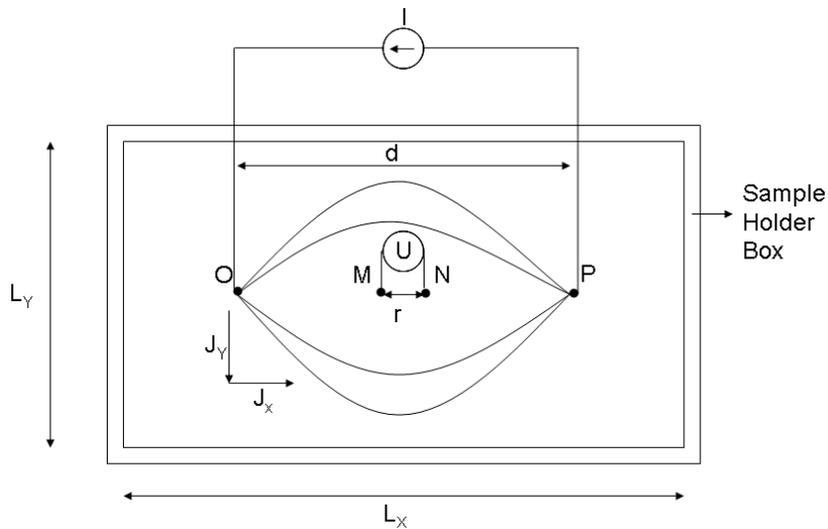

**Figure 6.** Top view of setup arrangement of the four-electrode measurement with line-contacts in a box of insulator.

However, at field measurements, see Figure 7 and 8, it is supposed that the distance (*d*) between the current electrodes is much less than their length $L_z$ in order to provide a dominantly horizontal current direction. Because the soil sample is "bottomless", vertical current lines still arise downward from the tip of the electrodes thus the length of the voltage electrodes is supposed to be much shorter than the length of the current electrodes. Then, the voltage electrodes are dominantly exposed to horizontal current density lines, and go along vertically aligned equipotential surfaces. Furthermore, the distance between the voltage electrodes is assumed to be much less than their length so that they probe only horizontal conductivity components and do no mix in other components due to adjoint current contributions, see [12, 13].

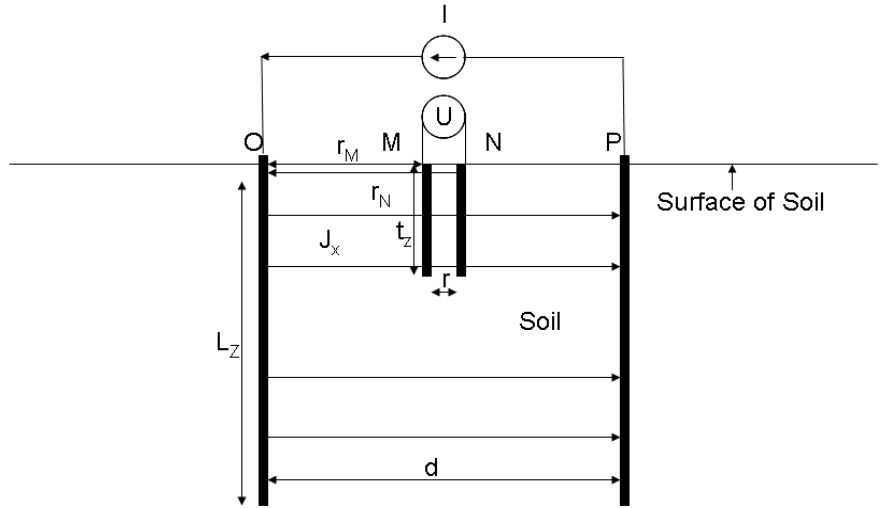

**Figure 7.** Side view of setup arrangement of the four-electrode measurement with line-contacts in field conditions.

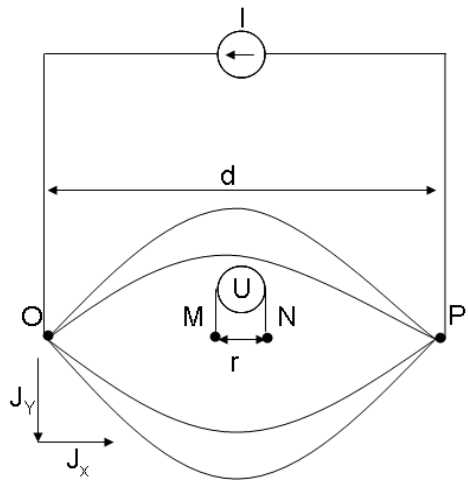

**Figure 8.** Top view of setup arrangement of the four-electrode measurement with line-contacts in field conditions.

It is easy to determine the normalized *VICOF* response because Equation 23 is valid here, too. However, due to the horizontal alignments of current density lines and vertical alignments of equipotential surfaces, it is straightforward to derive the exact value of the soil resistivity.

*Introducing compression VICOF and the phenomenological theory of compression and acceleration VICOF.*

By supposing that the current electrode is as long as the height of the soil cube $L_z$ and assuming that $d >> r = \overline{MN}$, $r >> r_E$, $\overline{OM} = \overline{NP} \cong \dfrac{d}{2}$, where $r_E$ is the radius of electrodes.

The assumptions about electrode lengths and distance considered above can be summarized as follows. In the box measurement set, the voltage electrode length $t_z >> r$.

For field measurement set, the current electrode length $L_z >> d$ must be satisfied.

The analysis of the scheme is as follows. Similarly to point contact measurement in 3-1, the differences are that the current enters through a single line-contact with radius $r_E$ and the other electrode (ground) is an infinitely large cylinder centered on the line-contact. Using the salami method between two cylinders of radius $r$ and $r + dr$ centered on the line electrode, the resistance contribution of this infinitesimally thin layer (of thickness $dr$) to the total resistance of the line-contact is given as:

$$dR = \frac{\rho \cdot dr}{2\pi r L_z} \quad . \tag{27}$$

The voltage is

$$dU = \frac{I\rho \cdot dr}{2\pi r L_z} \quad . \tag{28}$$

Considering only one electrode $O$, and then the voltage between $M$ and $N$ is

$$U_{MN,O} = \int_{r_M}^{r_N} \frac{I\rho}{2\pi r L_z} dr = \frac{I\rho}{2\pi L_z} \ln\left(\frac{r_N}{r_M}\right) \quad . \tag{29}$$

Similarly, considering only the other electrode $P$ and the opposite sign of electrical current there, the voltage between $N$ and $M$ induced by electrode $P$ is

$$U_{MN,P} = -\int_{d-r_M}^{d-r_N} \frac{I\rho}{2\pi r L_z} dr = \frac{I\rho}{2\pi L_z} \ln\left(\frac{d-r_M}{d-r_N}\right) \quad . \tag{30}$$

As a consequence, the total voltage between electrodes $M$ and $N$ is the superposition of $U_{MN,O}$ and $U_{MN,P}$

$$U_{MN} = \frac{I\rho}{2\pi L_z} \ln\left(\frac{(d-r_N)r_M}{(d-r_M)r_N}\right) \quad . \tag{31}$$

Thus using Equation 31, we can determine the resistivity from current/voltage measurement data with the following equation

$$\rho = \frac{2\pi L_z U_{MN}}{I \cdot \ln\left(\frac{(d-r_N)r_M}{(d-r_M)r_N}\right)} \quad . \tag{32}$$

For the normalized *VICOF* signal, we have the very same relation as for Equation 27:

$$\frac{\Delta \rho_v}{\rho} = \frac{2U_{MN}(f_c)}{U_{MN}(f_1)} \quad . \tag{33}$$

### 3.3. *Four-electrode measurement with plate-contacts*

The side view is the same as that of the four-electrode arrangement, see Figure 9, and for the top view see Figure 10.

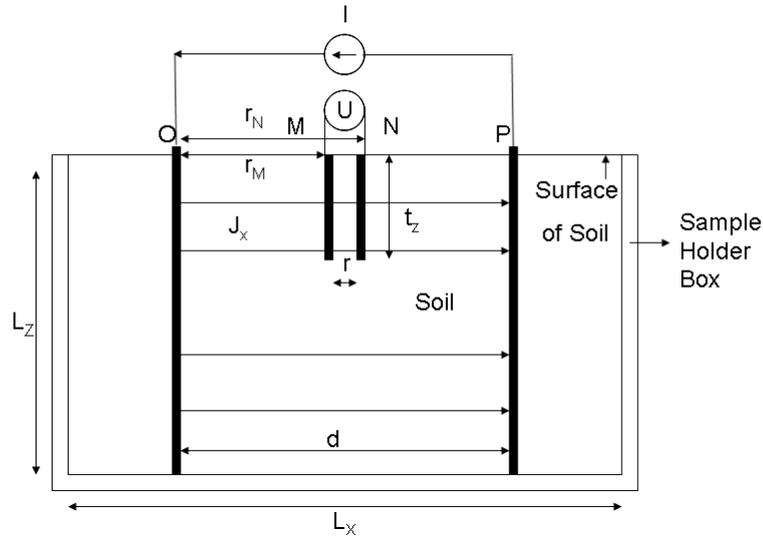

**Figure 9.** Side view of setup arrangement of the four-electrode measurement with plate-contacts in a box of insulator.

Suppose the electrodes have widths $L_Y$ and $L_z$. By assuming, $r \gg t_x$ and assuming $t_y, t_z \gg r$ we have approximately homogeneous current density distribution [12, 13] which are parallel horizontal lines. For field measurement set, assuming extra conditions that electrode length $L_z, L_y \gg d$ are needed to have homogeneous current density distribution.

*Introducing compression VICOF and the phenomenological theory of compression and acceleration VICOF.*

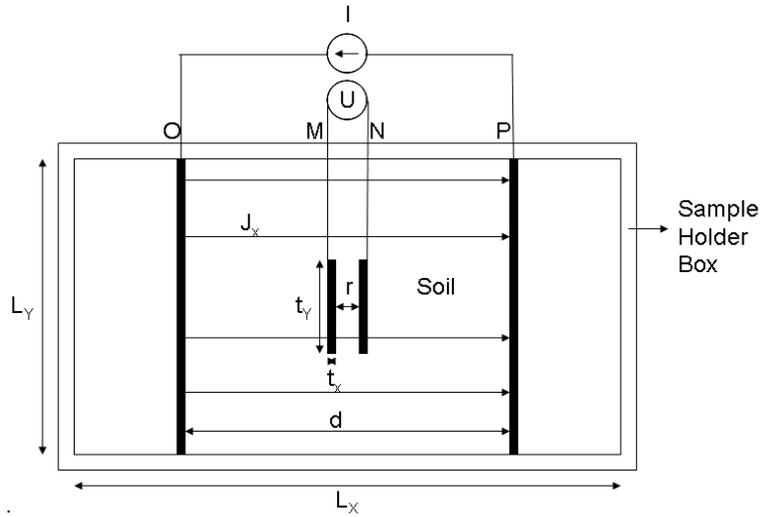

.

**Figure 10.** Top view of setup arrangement of the four-electrode measurement with plate-contacts in a box of insulator.

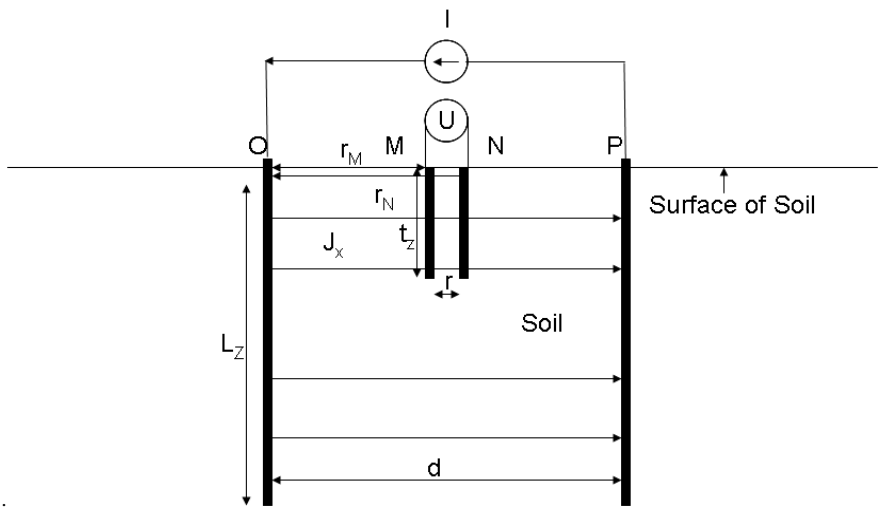

.

**Figure 11.** Side view of setup arrangement of the four-electrode measurement with plate-contacts in field conditions.

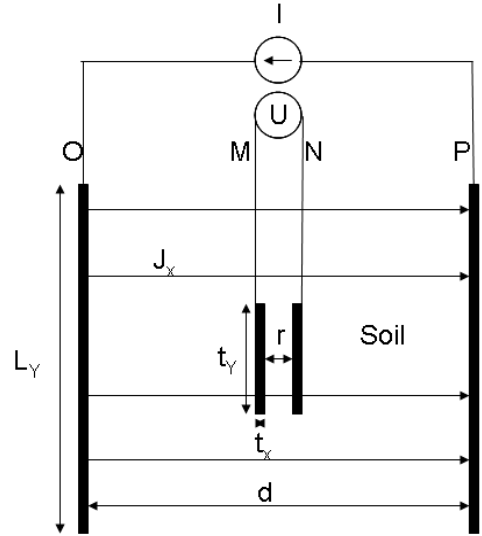

**Figure 12.** Top view of setup arrangement of the four-electrode measurement with plate-contacts at field conditions.

The resistivity of the soil is given as [11]:

$$\rho = \frac{AU_{MN}}{rI} \quad . \tag{34}$$

The cross-section area A is

$$A = L_Y \cdot L_Z \quad . \tag{35}$$

By substituting (34) into (35), we get the equation (36)

$$\rho = \frac{L_Y L_Z U_{MN}}{rI}(\Omega \cdot m) \quad . \tag{36}$$

And the relationship between normalized resistivity fluctuation and measured voltage is the same as equation (23) obtained from 3.1.

As shown in Figure 5 and 7, the applied force is perpendicular to the current in four-electrode measurement with line or plate-contacts. Pure transversal response can be produced with these electrode arrangements. However, the plate-contact setup is not easy to establish. As a consequence, four-electrode measurement with line-contacts is the most appropriate arrangement for *VICOF* with vertical compression and vibration methods.

*Introducing compression VICOF and the phenomenological theory of compression and acceleration VICOF.*

## 4. New schemes: vertical compression and vertical acceleration method

### 4.1. *Vertical compression method*

The vertical compression *VICOF* model is shown in Figures 13 and 14, which can be used in both laboratory and field measurements [14]. In laboratory, the soil cube is in a firm box and a pressure is applied by a heavy insulating top cover of mass *M* and a vibrator fixed to it, see Figure 13. In field conditions, the soil cube maybe confined by walls and a pressure is applied by vertical compression in *z* direction as shown in Figure 14 or, if not, the electrodes must be much shorter than the sizes of the top cover in order to provide homogeneous force conditions.

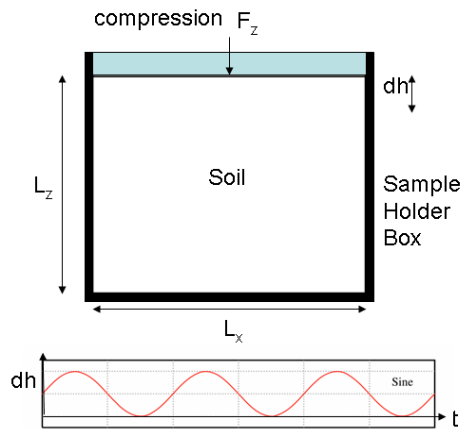

**Figure 13.** Vertical compression *VICOF* model in a box of insulator

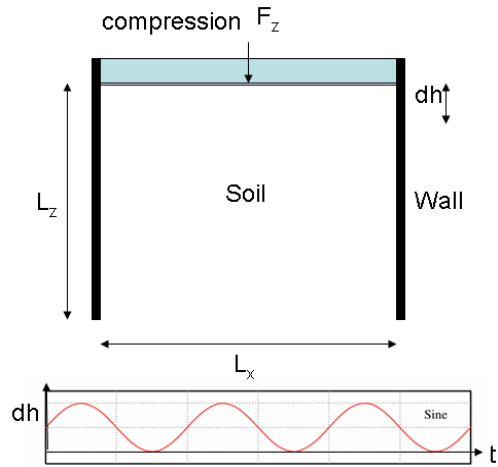

**Figure 14.** Vertical compression *VICOF* model in field conditions

For shallow boxes, we can neglect density changes coming from the weight of soil and ensure the applied force affect the same in different depths of the soil. Then the stress of the soil can be expressed as the equation below where $F_z$ is the applied periodic pressure.

$$\sigma_z = \frac{Mg + F_0 \sin \omega t}{L_x L_y} \;\;, \tag{37}$$

where the weight of the top cover must satisfy $Mg > F_0$ to provide a positive pressure at all times; here $g$ is the gravitational acceleration. We can write:

$$\sigma_z = \sigma_{z0} + \sigma_z(t) \;\;, \tag{38}$$

where $\sigma_{z0}$ is the DC stress and $\sigma_z(t) = F_0 \sin \omega t$ is the sinusoidal stress component generated by the vibrator.

Similarly to the horizontal acceleration-based vibration, the stress is transferred to the soil resistance and it generates periodic fluctuations of the soil resistance, see Equations 8-14.

### 4.2. *Vertical Acceleration Method*

The vertical acceleration *VICOF* model is shown in Figure 15, which is proposed for laboratory measurements [14]. The soil cube is in a firm box and applied by vertical vibration in *z* direction.

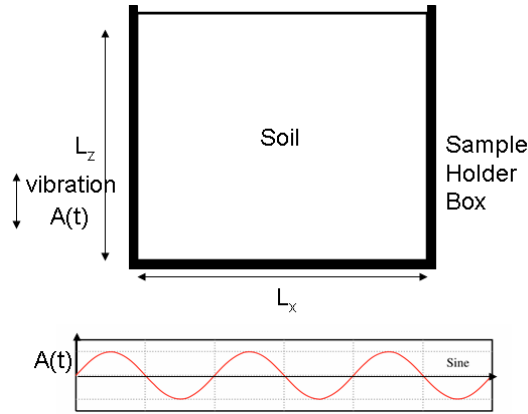

**Figure 15.** Vertical acceleration *VICOF* model

Suppose, we apply a periodic horizontal vibration

$$A_z(t) = A_0 \sin \omega t \;\;. \tag{39}$$

*Introducing compression VICOF and the phenomenological theory of compression and acceleration VICOF.*

In the direction $z$, where $A_0$ is the vibration amplitude and $\omega$ is the angular frequency of the vibration. The acceleration of the soil can be expressed as

$$a_z(t) = -\omega^2 A_0 \sin \omega t \ . \tag{40}$$

where $g > \omega^2 A_0$ to keep soil in the box at all times.

Similarly to the horizontal acceleration-based vibration theory, the acceleration of the soil induces a stress $\sigma$ (force per unit area) and strain $\epsilon$ (relative deformation) in the soil

$$\sigma_z(t) = \rho \cdot \omega^2 A_0 \ |\sin \omega t| \cdot f_\sigma(L) \ , \tag{41}$$

$$\varepsilon_z(t) = \rho \cdot \omega^2 A_0 \ |\sin \omega t| \cdot f_\varepsilon(L) \ . \tag{42}$$

Similarly to the horizontal acceleration-based vibration, the stress is transferred to the soil resistance and it generates periodic fluctuations of the soil resistance, see Equations 8-14.

The advantage of the method is that the acceleration is aligned with the electrode axis thus lose electrodes have a smaller impact on the measurement accuracy/reproducibility.

## 5. Summary

We have established a complete *VICOF* model of electromechanical response, which describes the relationship between applied vibrations, stress and strain, resistance and resistance variation. We also proposed new measurement schemes, such as vertical compression VICOF with four line- or plate-contact electrodes for laboratory and field conditions, and vertical acceleration VICOF with four line- or plate-contact electrodes for laboratory conditions. It is expected that using vertical compression or vertical acceleration VICOF methods will reduce the impact of lose electrode contact on measurement accuracy.